\documentclass[conference]{IEEEtran}
\usepackage{cite}
\usepackage{caption}
\usepackage{amsmath, amssymb}
\usepackage{graphicx}
\usepackage{hyperref}
\usepackage{url}
\usepackage[table]{xcolor}
\usepackage{booktabs}
\usepackage{multirow}
\usepackage{enumitem}
\usepackage{listings}
\usepackage{subcaption}
\usepackage{float}
\usepackage{algorithm}
\usepackage{algpseudocode}
\usepackage{microtype}   

\usepackage{longtable}
\hypersetup{
    colorlinks=true,
    urlcolor=blue,
    citecolor=blue,
    linkcolor=blue
}
\usepackage{subcaption}
\usepackage{soul}

\lstdefinelanguage{Gherkin}{
  morekeywords={Feature,Scenario,Given,When,Then,And,But},
  sensitive=false,
  morecomment=[l]{\#},
  morestring=[b]",
}
\usepackage{comment}
\lstdefinelanguage{json}{
  morestring=[b]",
  showstringspaces=false,
}
\title{
Multi-Agent Specification-based Metamorphic Testing of FMU-Based Simulations
}
\author{Anonymous Authors}
\author{
\IEEEauthorblockN{%
Ashir Kulshreshtha\IEEEauthorrefmark{1},
Abdullah Mughees\IEEEauthorrefmark{1},
Gaadha Sudheerbabu\IEEEauthorrefmark{1},
Tanwir Ahmad\IEEEauthorrefmark{1},
Kristian Klemets\IEEEauthorrefmark{2}}
Dragos Truscan\IEEEauthorrefmark{1}, and
Mikael Manng{\aa}rd\IEEEauthorrefmark{3},
\IEEEauthorblockA{\IEEEauthorrefmark{1}\AA bo Akademi University, Finland
\texttt{(firstname.lastname@abo.fi)}}
\IEEEauthorblockA{\IEEEauthorrefmark{2}University of Turku, Finland
\texttt{(firstname.lastname@utu.fi)}}
\IEEEauthorblockA{\IEEEauthorrefmark{3}Novia University of Applied Sciences, Finland 
\texttt{(firstname.lastname@novia.fi)}}
}

\begin{document}

\maketitle

\begin{abstract}
In many industrial domains, the Functional Mock-up Interface (FMI) is used to exchange simulation models as Functional Mock-up Units (FMUs) across different partners using various modelling tools. This opens up the possibilities for simulation-based verification and validation using FMUs for ensuring reliable system behaviour. However, deriving effective test oracles for these simulation models remains challenging due to the absence of explicit expected outputs. This limits the applicability of conventional testing approaches, which require access to the internal workings of the systems. Metamorphic testing (MT) addresses this limitation by leveraging metamorphic relations (MRs), but extracting such relations from specifications remains largely a manual and error-prone process.
To address this challenge, we propose an LLM-powered multi-agent workflow for specification-based metamorphic testing of FMU-based simulation models. The approach takes functional and interface specifications as input and orchestrates multiple agents to extract requirements and derive MRs. These MRs are expressed using Given–When–Then patterns to structure input conditions (Given), transformations (When), and expected output behaviours (Then). These relations are then used to generate metamorphic test cases, execute simulations, and evaluate output consistency across multiple sessions. We evaluate the approach on a Lube Oil Cooling system FMU, demonstrating its ability to automatically generate meaningful MRs and corresponding test cases. 
Preliminary results indicate that the proposed workflow can effectively support the systematic verification and validation of dynamic simulation models by reducing manual effort and improving test generation.
\end{abstract}

\begin{IEEEkeywords}
Metamorphic testing, multi-agent systems
\end{IEEEkeywords}

\section{Introduction}\label{sec:introduction}

In modern industrial systems, model-based design is an increasingly integral approach to system development, in which components are designed and validated through simulation prior to physical implementation \cite{cederbladh2024early}. Organizations typically use various specialized modelling and simulation tools for model-based development and often collaborate with internal and external partners in this process. Systems development strategy using component-based design in such collaborative settings necessitates a standardized way of model exchange and co-simulation. 

The Functional Mock-up Interface (FMI) \cite{blochwitz2012functional} is a standard for tool-independent exchange of dynamic simulation models. It defines a container and an interface to exchange simulation models as Functional Mock-up Units (FMUs). An FMU is a ZIP archive that typically contains: (i) binaries and/or source to execute the model, (ii) model description, and (iii) optional resources (such as documentation files, maps and tables needed by the model, and/or all object libraries or DLLs that are utilized).

This encapsulation brings along several challenges for validation and verification of the FMU components: a) often the source code of the model and its internal functionalities are not included in the package with it, allowing only for specification-based testing techniques to be applied, and b) the test oracles are not explicitly defined since typically the testing of such simulation models is done manually by domain experts. 

However, the simulation model of a component/system is designed based on the specifications of the real system it represents and is available for validation early in the development. This opens up the possibility of early simulation-based testing using specification-based testing techniques.

Metamorphic testing (MT) is a specification-based testing technique that can be used to test systems lacking explicit test oracles~\cite{chen2020metamorphic}. MT checks whether multiple executions of the system under test (SUT) satisfy specific necessary properties, called \textit{metamorphic relations} (MRs). It starts with a \textit{seed input} and derives one or more \textit{follow-up inputs} by applying a \textit{metamorphic transformation}. Instead of checking one output against a fixed expected value, the test verdict is assigned based on whether the MR that links the seed and follow-up outcomes holds 
\cite{liu2012new}. MT has been applied in many application domains, including simulation and modelling ~\cite{segura2018metamorphic, nunez2015methodology, lindvall2017metamorphic, olsen2018increasing, sudheerbabu2025validation}.

Despite prior studies on generating MRs and MT for simulation models, there is still no FMU-focused approach that starts from functional and interface specifications and continues through MR generation, test generation, simulation execution, and test-quality assessment. 
Moreover, identifying suitable MRs remains a practical challenge when exploring the applicability of metamorphic testing to new application domains \cite{Segura2016Survey, chen2018metamorphic}. 
FMU-based simulations introduce practical constraints, for example, the internal model implementation may be unavailable, the executable interface is limited to exposed FMU variables, and not every requirement in the functional specification can be directly converted into a valid input-output MR.
These constraints motivate the need for an automated workflow that identifies MRs from specifications, then designs, executes metamorphic tests, and reports requirement coverage for test adequacy and mutation score for test quality.

Large language models (LLMs) are increasingly used for automated test generation, and recent studies~\cite{luu2023chatgpt, zhang2023automated, shin2024generatingexecutablemetamorphicrelations} show that LLMs can identify MRs in different application domains, such as web applications, autonomous driving systems, and embedded systems. Although these studies demonstrate the effectiveness of LLMs in generating innovative and diverse MRs, they also emphasize the need for a domain expert to validate the MRs to ensure a high accuracy. Open challenges remain in systematically identifying effective MRs and reducing reliance on domain experts~\cite{chen2018metamorphic}.

In this paper, we investigate the following research questions: 
\begin{itemize}[leftmargin=*]
   \item \textit{RQ1:}  Can a multi-agent workflow systematically perform automated MT from the functional and interface specification of an FMU?
  \item \textit{RQ2:} To what extent does the approach extract specification-linked MRs and executable metamorphic test cases based on the FMU specification?

   \item \textit{RQ3:} What is the quality of the generated test cases?  

\item \textit{RQ4:} What are the main runtime costs of the workflow across extraction, MR generation/refinement, test generation, and test execution?
   
\end{itemize}

To answer these research questions, we introduce \mbox{\textbf{AgenticMeta}}, an LLM-powered multi-agent workflow to support metamorphic testing of FMU-based simulation models using the functional and interface specifications as the primary input. The key contributions of this paper are as follows:
\begin{itemize}[leftmargin=*]
    \item A multi-agent approach for FMU-based metamorphic testing that extracts requirements from functional and interface specifications and then identifies and selects MRs using LLM-powered agents; 
    \item The extracted MRs are specified as GWT requirement patterns, allowing for better processing by LLMs to generate metamorphic tests with customized input transformations and output relations for simulation models;

    \item A feasibility evaluation on a Lubricating Oil Cooling system FMU across multiple independent sessions, reporting requirement coverage, test execution outcomes, runtime statistics, and output-level mutation analysis.
  \end{itemize}

We will evaluate the approach on a dynamic simulation model, a simplified version of a Lubricating Oil Cooling (LOC) system \cite{LOC_NoviaRDISeafaring2024}, packed as an FMU. The LOC system models a lube oil cooler that transfers the heat from the lubrication oil to the cooling water circuit of a marine engine unit.

The rest of this paper is organized as follows. Section~II discusses related work. Section~III presents the AgenticMeta workflow and its main components. Section~IV describes the experimental design and reports the evaluation results. Section~V discusses the limitations. Section~VI concludes the paper and outlines future work.


\section{Related Work}
\label{sec:related_work}

\subsection{LLM-assisted MR Generation}

The application of LLMs for automating different phases of MT has been investigated. Several studies exploring the applicability of ChatGPT for automatically generating metamorphic relations demonstrated its usability \cite{zhang2023automated, luu2023chatgpt, shin2024generatingexecutablemetamorphicrelations}. In \cite{luu2023chatgpt}, a zero-shot prompting strategy is used to generate MRs and these MRs are verified by domain experts. Their study, conducted across nine software systems, suggested that ChatGPT-generated MRs can be used to generate innovative MRs and highlighted the need for human-in-the-loop validation. ChatGPT is used to generate MRs for autonomous driving systems  \cite{zhang2023automated}, and it is reported that the approach can generate diverse MRs that effectively improve coverage using a feedback-based prompting strategy. The third study \cite{shin2024generatingexecutablemetamorphicrelations} uses a few-shot prompting strategy to derive MRs from natural-language requirements and convert them into executable forms using a domain-specific language called SMRL. Their work shows that LLMs can help identify transformation-based test relations and reduce manual effort.

\subsection{LLM-assisted Metamorphic Testing}
LLM-assisted metamorphic testing has also been applied to embedded systems and the simulation and modelling domain. More recent work ~\cite{automt2025} presents a multi-agent tool that applies metamorphic testing to autonomous driving systems. It derives MRs from driving rules, creates follow-up tests from existing tests, executes them, and reports violations. In contrast, our approach targets FMU-based dynamic simulation models.  Our approach differs mainly in how it infers MRs, structures tests, and refines them. We also address practical instability in large language model outputs by separating generation and refinement. A stable model produces initial MRs, and a stronger model refines the results to better match the specification and reduce unsafe patterns. This can improve repeatability, but full determinism is still difficult in LLM-based pipelines \cite{atil2024non}.

Compared with prior MT work for simulation models, AgenticMeta focuses specifically on FMU-based simulations where the testable interface is constrained by the variables exposed in the FMU model description. 
Earlier GWT-based MT work for dynamic simulation models \cite{sudheerbabu2025validation} provides a useful structure for expressing MRs; this paper extends that work by automating requirement extraction, MR generation, MR refinement, test generation, and validation through LLM-based agents.
Recent LLM-assisted MR-generation studies mainly focus on producing candidate MRs, whereas AgenticMeta automates the entire MT process, integrating MR refinement, test execution, and output-level mutation analysis into a single process.
Compared with multi-agent MT for autonomous driving systems, our target domain is FMU-based dynamic simulation, where inputs and outputs are continuous time-series variables and where valid MRs must conform to the FMU interface constraints. Therefore, the primary contribution of this work resides not in the application of MT or LLMs in isolation, but in the integration of these components into a traceable, FMU-oriented MT workflow.

\section{Overview of the Approach}
\label{sec:approach}

\subsection{Conceptual approach}

\begin{figure}[h!]
\centering
\includegraphics[trim={0cm 0.1cm 0cm 0.1cm},clip, width=8.7cm]{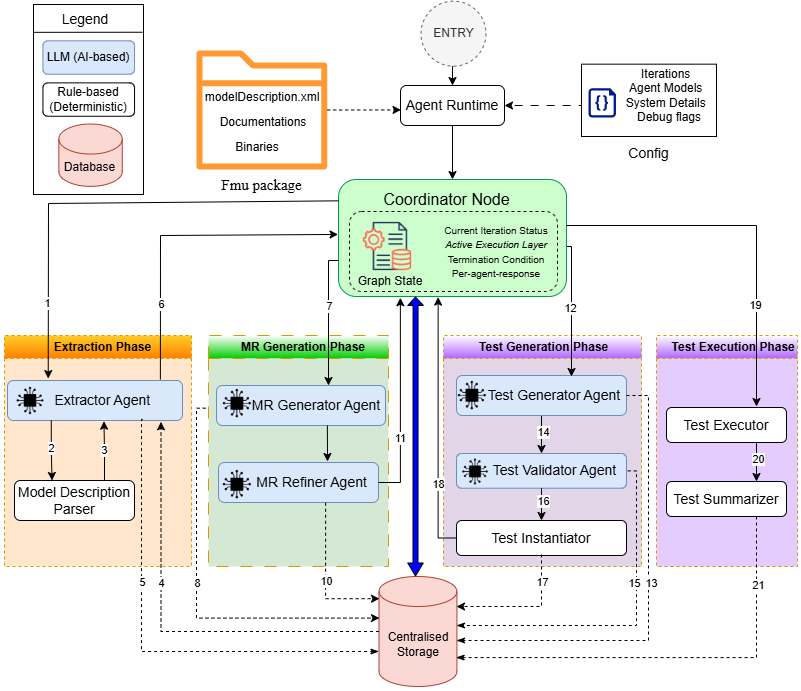}
  \caption{Overview of the Multi-Agent workflow}
  \label{fig:agentic_architecture}
\end{figure}

Our approach aims to augment the main steps of the metamorphic testing process, namely, systematic identification and selection of MRs from functional and interface specifications, and metamorphic test generation in patterns similar to Gherkin's language Given-When-Then (GWT) requirement patterns~\cite{wynne2012cucumber} with an LLM-powered multi-agent workflow for validation of dynamic simulation models. The workflow comprises four phases: extraction, MR generation, test generation, and test execution, as illustrated in Figure \ref{fig:agentic_architecture}. The first three phases are performed by specialized agents, whereas the test execution phase is performed by deterministic custom tools. The overall process is coordinated by a Coordinator node that orchestrates interactions between agent-based and deterministic components.

The approach takes as input the simulation model, packed as an FMI-standard compliant FMU, which comprises the functional specification, model description and execution binaries to facilitate the metamorphic testing. In the first step of our approach, we extract system properties to be selected for MR selection from the functional and interface specifications of the simulation model. The system properties specified in the functional specification are extracted as test conditions. In addition to this, the cause-and-effect relationship between the input and output variables in a specified property is extracted from the specification document. From the model description XML file in the FMU package, the model variables and the interface specification, including inputs, outputs, parameters, and their details (name, description, variability, unit, datatype, minimum, and maximum values), are extracted. The extraction of relevant information and storing it in such a structured manner as specified above streamlines the identification and selection of the MRs in the next phase. 

Subsequently, in the MR generation step, MRs are generated with an MR identifier and are mapped to the test condition identifier it corresponds to, along with the test scenario description. The generated MR has the metamorphic source and follow-up test specified in the GWT pattern with \textit{Given}, \textit{When}, and \textit{Then} keywords. The \textit{Given} states the initial conditions to generate seed inputs for the source test execution, \textit{When} states the metamorphic input relation for transforming the seed inputs to morphed inputs, and \textit{Then} step states the acceptance criteria as a metamorphic output relation between the outputs generated using the seed inputs and follow-up inputs. In the test generation step, the test design specified in each MR is converted into seed and follow-up input signals as a time series, each over a specified time interval for each SUT input. In the test execution step, which is the last step in the workflow, we run and evaluate the generated time series against the FMU system. 

\subsection{Multi-agent workflow}
\label{sec:agentic_workflow}

The agents, central to our workflow, perform well-defined tasks in a coordinated manner to facilitate LLM-assisted metamorphic testing. The approach follows a 'hub and spoke' architecture where the pipeline is controlled by a \textit{Coordinator node} which acts as the hub monitoring, orchestrating the workflow and functioning of the agents with clearly defined roles. The approach is iterative and incremental. 

The inputs to the workflow are: the FMU file of the system under test, including the model description, functional specification as pdf and binaries of the model, and a set of configuration parameters for the workflow, as follows: 

The following parameters can be used to customize the workflow for a given case study: 
    \begin{itemize}[leftmargin=*]
    \item \textit{system\_name}: Name of SUT.
    \item \textit{system\_abv}: Short abbreviation for the SUT.
    \item \textit{fmu\_path}: Path to the FMU file used for simulation.
    \item \textit{output\_dir}: Directory name where run outputs are saved.
    \item \textit{max\_iterations}: Max number of iterations per run.
    \item \textit{mr\_count}: Max number of MRs to generate per iteration.
    \item \textit{test\_cases\_per\_mr}: Max number of test cases to generate per MR.
    \item \textit{llm\_provider}: LLM service provider identifier.

    \end{itemize}

\subsubsection{Coordinator Node}
The \textit{Coordinator node} maintains the global execution state, determines the phase of the workflow, and routes flow to the corresponding node. This design separates control-flow management from task-specific execution, enabling each LLM agent or deterministic module to operate on a well-defined responsibility while the Coordinator governs the overall progression of the pipeline. A shared graph state is a central context that manages execution information, tracks intermediate states, and propagates coordination signals across nodes throughout the workflow.

At each workflow step, the Coordinator inspects the \emph{phase} field in the shared graph state and applies a phase-to-node mapping to select the next executable component. After an iteration is completed, it records execution-related statistics. If the configured maximum number of iterations has not yet been reached, the Coordinator prepares for the next iteration by initialising iteration-specific directories and resetting iteration local artifacts to their initial state. When the iteration count is exhausted, the Coordinator marks the workflow as completed and exits the execution according to the configured stopping condition.

\subsubsection{Extractor Agent} 

\begin{figure}
  \centering
  \includegraphics[trim={0cm 0.95cm 0cm 1cm},clip, width=6cm]{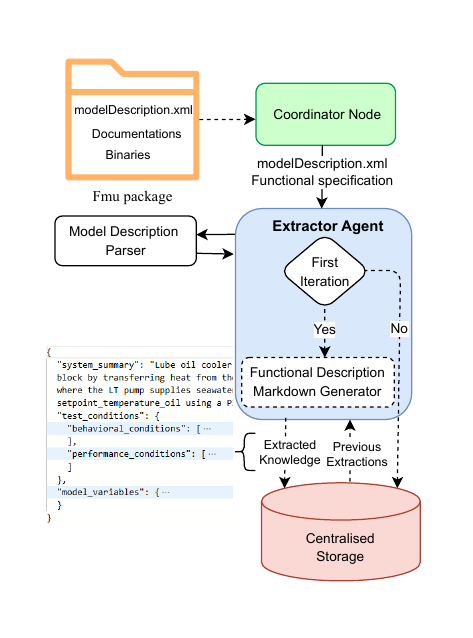}
  \caption{Workflow of Extractor Agent}
  \label{fig:extraction_output}
\end{figure}

The \textit{Extractor Agent} operates by transforming the input functional specification document through a structured conversion pipeline, then formulating an output in an LLM-friendly content format (see  Figure~\ref{fig:extraction_output}). It transforms the functional specification document from PDF to Markdown, a structured format that enhances the processing and accuracy of agents that accept it as input for proceeding with further steps of MT. The system prompt to define the role of the agent for performing the information extraction is shown in Listing~\ref{tab:System_prompt_extractor}.

\captionsetup[table]{name=Listing} 

\begin{table}[h]
\begin{center}
\begin{tabular}{|p{0.9\linewidth}|}
\hline
\rowcolor{gray!30}
\textbf{System Prompt} \\
        Role:\\
        You are a domain expert in Functional Mockup Interface (FMI) and Functional Mockup Unit (FMU) systems, with strong experience in requirement extraction and structured evidence preservation.\\
        Core Principles:\\
        1. Preserve document context \\
        2. Accept equivalent table layouts (grid/pipe/html/text) as the same semantic structure.\\
        3. Document is authoritative — prioritize document facts as the single source of truth. \\
        4. Verbatim evidence — every extracted fact (relationship, formula, constraint, test condition) must be grounded in a direct quote or precise reference from the document.\\
        5. Tables are primary sources — treat input/output tables, parameter tables, and model description tables as the authoritative metadata source for variable specifications.\\
        6. Precision over inference \\
        6. Preserve relationships — for causal links and behavioral rules, write actionable statements that capture the trigger, condition, direction of change, and operational consequence.\\
        7. Valid JSON output only — return well-formed JSON without markdown fences, commentary, or embedded natural language.\\

\hline
\end{tabular}
\end{center}
\caption{System prompt of Extractor Agent}
\label{tab:System_prompt_extractor}
\end{table}

The model interface data and the constraints for input and output variables, including names, units, causality, and allowed ranges, are extracted from the \texttt{modelDescription.xml} file of the FMU. 
A custom XML parser script extracts and converts the information to JSON format.

To summarize, the \textit{Extractor Agent} generates extraction output in a structured schema as defined in the task prompt. The output in nested JSON format comprises of the following information to facilitate the MR generation: system summary test conditions (categorized by property-type if relevant details are present in the functional specification), and model variables. 

\subsubsection{MR Generator Agent} 

In each iteration, the \textit{MR Generator Agent} proposes a number of MRs as specified by the \textit{mr\_count} workflow configuration parameter. The task prompt designed for the \textit{MR Generator Agent} guides the generation of candidate MRs. An excerpt of the task prompt, which constraints the generated MRs to follow the schema and to assign priorities as per the rules, is shown in Listing \ref{tab:task_prompt_MR_generator}.

\begin{table}[h]
\begin{center}
\begin{tabular}{|p{0.9\linewidth}|}
\hline
\rowcolor{gray!30}
\textbf{Task Prompt} \\
        Task:\\
    - Generate up to {mr\_count} NEW, UNIQUE, CONCRETE and test conditions focused Metamorphic Relations (MRs).\\
    - Use concise test conditions focused **Given-When-Then** form.\\
    - Each MR MUST include: id, req\_ids, scenario, given, when, then.\\
    - Prevent duplicates with current \& previous MRs.\\
    - Keep the wording compact and clear in given, when and then.  \\  
    Prioritise MR generation categories in this order:\\
    {priority}   \\
    Constraint by Category:\\
    - Behavioral\\
    - Performance\\

\hline
\end{tabular}
\end{center}
\caption{Task Prompt of MR Generator Agent}
\label{tab:task_prompt_MR_generator}
\end{table}

The prompt explicitly requires the candidates to be different from those generated in previous iterations by providing the agent with the accumulated MR history of up to three previous generations. Each candidate MR is assigned a unique MR identifier and linked to the requirements it addresses by listing the test condition identifier and variable relationship identifier. 

\begin{figure}[h]
  \centering
  \includegraphics[trim={0cm 0.65cm 0cm 0.7cm},clip, width=6cm]{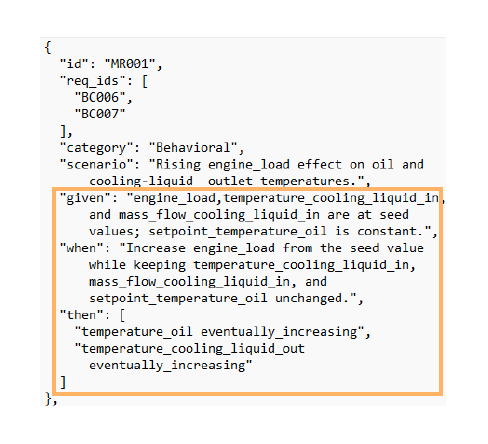}
  \caption{MR generated by the agent in Given-When-Then format}
  \label{fig:example_mr}
\end{figure}

The agent generates MRs as GWT patterns under a strict JSON schema using the rules defined in the task prompt, so that subsequent stages can reliably generate metamorphic tests. Figure~\ref{fig:example_mr} shows an example MR generated by the agent in the structured schema as per the rules defined in the task prompt in Listing~\ref{tab:task_prompt_MR_generator}. In the example, \textit{MR001}, the initial condition values for the seed input signals are stated in the \textit{Given} part. The metamorphic transformation in the \textit{When} part of the inferred MR states that the input signal \texttt{engine\_load} transforms from its seed value to a morphed value using the relational operator \textit{'increase'}. The agent generates transformations aligned by the rules specified in its task prompt for the \textit{When} using patterns such as \textit{'STEP', 'RAMP'}, considering the scenario and system properties to be validated. 

The metamorphic output relation to determine the test outcome is stated in the \textit{Then} part using comparative and temporal operators such as the ones listed below:

\begin{itemize}[leftmargin=*]
\itemsep0em
\item \textit{Eventually\_Increases than}: Implies that morphed outputs of the selected output variable should eventually increase than the corresponding seed outputs.
\item \textit{Eventually\_Decreases than}: Implies that morphed outputs of the selected output variable should eventually decrease than the corresponding seed outputs.
\item \textit{Proportional\_to}: Implies that morphed outputs of the selected output variable should always be proportional to the corresponding seed outputs.
\item \textit{Equal\_to}: Implies that morphed outputs of the selected output variable should always be equal to the corresponding seed outputs.
\item \textit{Settles\_within}: Implies that seed and morphed outputs of the selected output variable should settle within X seconds to a defined set\_point value.
\end{itemize}

In total, the agent generates $mr\_count * iteration\_count$ number of candidate MRs, where $mr\_count$ is defined by \textit{MR Generator Agent} and  $iteration\_count$ by the \textit{Coordinator node} for metamorphic test generation.  These generated MRs will be subsequently passed to the next agent for further validation before proceeding to the test generation.  

\subsubsection{MR Refiner Agent} validates if the system property specified as a test condition from the functional specification is correctly captured and translated into MRs by the \textit{MR Generator Agent}. The validation criteria for each MR take into consideration the factual correctness, category consistency, constraint compliance, causal validity, and testability.

Any missing detail or incorrect information in each MR is verified and refined using refinement rules by the \textit{MR Generator Agent}. In the output schema of the refined MRs, a field named \textit{feedback} provides the information on the refinement performed for each MR, in order to enhance the explainability of the decision, and the field named \textit{dropped} with values as either \textit{'true'/'false'} indicates whether the particular MR should be accepted with refinement or discarded.

\begin{figure}[h]
  \centering
  \includegraphics[trim={0cm 0.8cm 0cm 0.8cm},clip, width=6cm]{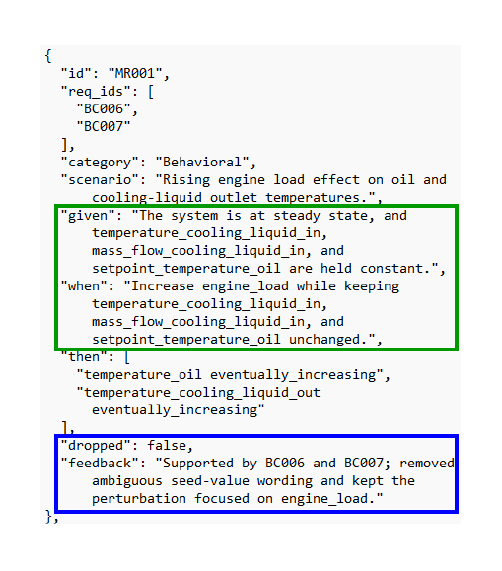}
  \caption{MR refined by the agent and the feedback provided}
  \label{fig:example_refined_mr}
\end{figure}

The repair loop is limited to a configurable number of attempts as defined in the pipeline configuration of the \textit{Coordinator node}. The refined version of the example MR shown in Figure~\ref{fig:example_mr} is presented in Figure~\ref{fig:example_refined_mr}, where the refined part is highlighted in green, and the \textit{feedback} on the refinement performed is highlighted in blue.

\subsubsection{Test Generator Agent} 

 The \textit{Test Generator Agent} generates a configurable count of metamorphic tests for each MR received for the previous phase. In this step, the simulation parameters of the model, such as the start and end times, and the system I/O constraints, are considered. The seed input values are assigned based on the initial conditions in the extraction output. The concrete values for performing the seed to follow-up transformation of input signals are defined in this step using a sampling algorithm. The algorithm uses the initial condition values extracted from the functional specification as the baseline for generating follow-up inputs using the input relation in the \textit{When} part of MR. 
 
 The test design also accounts for the rule that any step/ramp change to any input signal should be placed near the start of the simulation time window ($\sim$ 10-25\%). The rules also reinforce that any set-point values among the input signals should remain unchanged. Each MR/refined MR from the previous phase guides the generation of one or more tests. 

\subsubsection{Test Validator Agent} 

The \textit{Test Validator Agent} primarily validates the test inputs and MRs in the test case. It checks the test case against the functional specification and the metamorphic relation patterns used to generate the test inputs. The validation is performed based on repair rules such as data-type and boundary checks for test input values, type of metamorphic relation pattern, and simulation-specific time constraints.  In the output schema of validated tests, two fields named \textit{fixed} and \textit{dropped} capture the validation decision, and a summary is provided by the agent in the field \textit{validation\_summary}. The decision categorizes a test as any one of the following:

\begin{itemize}[leftmargin=*]
 \item If the test case is fully valid, keep it unchanged and set fixed=false, dropped=false.
 \item If the test case can be safely repaired, repair it, apply the smallest possible fix, and set fixed=true, dropped=false.
 \item If the test case is fundamentally broken or unsafe to repair, set fixed=false, dropped=true.
\end{itemize}
To summarize, the valid/repaired tests are passed to the test instantiator, which generates the input time series for all the input signals. The test input signals for which the metamorphic transformation is applied capture the evolution of values over time, and the other inputs are represented as constant vectors. These generated input signals are wrapped into an input object and proceed to the test execution phase. 

\subsection{Implementation}

AgenticMeta is implemented in Python as a layered workflow built around LangGraph \cite{LangGraph}. The implementation separates the pipeline into five layers: the orchestration layer, the specification and variable extraction layer, the MR generation and refinement layer, the test generation and validation layer, and the test execution layer. Each layer is integrated with a persistence mechanism that stores its inputs, outputs, intermediate results, and state, enabling traceability, debugging, and post-run evaluation across the complete workflow. This separation helps with debugging and evaluation, since intermediate artifacts are stored after each phase, the errors can be traced to a specific layer, and individual components can be modified or replaced without affecting the rest of the workflow. The layered design also allows different pipeline settings to be used at different stages, enabling isolated evaluation of each stage, more flexible reasoning early on, and more stable behaviour during execution. Overall, this structure makes AgenticMeta easier to extend and adapt to new tasks.

\section{Evaluation}
\label{sec:evaluation}

This section evaluates AgenticMeta in order to provide an answer to the research questions discussed in Section \ref{sec:introduction}.

\subsection{Case Study}

For evaluation, we use a simplified version of a Lubricating Oil Cooling (LOC) system.
The LOC system has a proportional–integral (PI) controlled valve that regulates the lubrication oil temperature at a constant set-point at the engine inlet. The controller aims to keep the lubrication oil temperature at the outlet within the specified boundary values under all operating conditions. The Lubricating Oil Cooling system has the following input variables, 
\texttt{temperature\_cooling\_liquid\_in}, \texttt{mass\_flow\_cooling\_liquid\_in}, \texttt{engine\_load}, and \texttt{setpoint\_temperature\_oil} (held constant throughout the simulation) and following output variables \texttt{temperature\_oil}, \texttt{position\_valve},
\texttt{temperature\_cooling\_liquid\_out}, and \texttt{mass\_flow\_cooling\_liquid\_out}.
The simulation model of the LoC system has been implemented and tested manually by the designers, and in the context of this evaluation, it is used as the ground truth.
The case study and tool were deployed locally on a Windows-based PC (13th Gen Intel Core i9-10900X, 3.70 GHz, 64~GB RAM, 4~GB dedicated graphics card, 2~TB storage). 
We use FMPy \cite{FMPy}, a Python library, to simulate the FMU and execute the tests generated by the workflow.

\subsection{Workflow configuration}
    
We used the following workflow configuration for our experiments. Some of these values were chosen based on empirical evaluations. All agents use OpenAI as the model provider and the GPT‑5.4‑mini model. Reasoning effort is set to none for the Extraction agent, low for test generation, medium for MR generation and test validator, and high for the MR refinement agent. The Extraction agent uses temperature 0 (temperature is only applicable with \emph{reasoning effort: none}). Each run consists of 5 iterations, with up to 5 MRs per iteration and 5 test cases per MR.

\subsection{Evaluation metrics}
\label{sec:experimental_design}

We evaluate AgenticMeta using the following metrics: 
\emph{Requirements coverage} denotes the number of requirements extracted from the specification and mapped into metamorphic test scenarios produced per test session. 
\emph{Test Case Summary} enumerating generated test cases, their MR assignments, and pass/fail outcomes, providing an indicator of workflow effectiveness.
\emph{Runtime statistics} to measure the efficiency of the workflow across different phases and units.

\emph{Mutation score} to measure the quality of metamorphic tests generated by the approach successfully executed (e.g., without any syntax errors) against the SUT. Since we do not have access to the internal specification of the FMU, we imitate possible design and implementation mistakes by applying systematic changes (\textit{mutations}) on the output signals. These mutations are defined based on a set of \textit{mutation operators} that are systematically applied, one at a time, to each output; this will result in a slightly incorrect version (\textit{mutant}) of the FMU 
In this work, we use the following mutation operators:

\begin{itemize}[leftmargin=*]
    \item \textit{Mirror Mutation}: Replaces an output signal with its mirror version from the start time step of the simulation horizon.
    \item \textit{Crossover Mutation}: Select two output time series of different metamorphic relations for a given testcase and pick a specific crossover site and interchange the values of the time series signals after the crossover site. 
    \item \textit{Polynomial Mutation}: Given a variable x in a time series in the range ($x_{min}$, $x_{max}$), we slightly change it by a random amount determined by a polynomial distribution.	
\end{itemize}

The process starts with the execution of the generated test cases on the original version of the FMU. To save on the simulation time, we apply the mutation operators directly on the recorded output time series for each output of the passed test cases. Each application of a mutation operand to an output time series will result in a new mutant. We compare the seed output and the mutated output using the output MRs and assign a pass/fail verdict. Finally, we calculate the test adequacy as a \textit{mutation score}, with values between 0 and 1, as the ratio between the number of mutants on which tests failed (killed mutants) and the total number of mutants created. 


\subsection{Results}
We evaluated the workflow across 10 independent sessions of the LOC case study. Each run used the same configuration.
\setcounter{table}{0}
\captionsetup[table]{name=TABLE}
\begin{table*}[t]
\centering
\caption{Runtime statistics across 10 sessions}

\label{tab:runtime_stats}
\footnotesize
\begin{tabular}{c|ccccc|ccc}
\toprule
\multirow{2}{*}{\textbf{Sessions}} &
\multicolumn{5}{c|}{\textbf{Phase level statistics (s)}} &
\multicolumn{3}{c}{\textbf{Unit level statistics (s)}} \\
\cmidrule(r){2-6} \cmidrule(l){7-9}
& Total Exec. & Extraction & MR Gen & Test Gen & Test Exec
& Gen. Time/TC & Gen. Time/MR & Exec. Time/TC \\
\midrule
1 & 771.39 & 81.74 & 540.28 & 110.24 
& 39.14 & 2.69 & 45.02 & 18.82 \\

2 & 913.54 & 78.70 & 617.23 & 160.21 
& 57.41 & 2.67 & 41.14 & 15.23 \\

3 & 881.11 & 81.04 & 619.95 & 132.72 
& 47.40 & 2.77 & 47.68 & 18.36 \\

4 & 1074.30 & 88.94 & 773.78 & 159.77 
& 51.81 & 2.90 & 48.36 & 19.53 \\

5 & 900.78 & 79.27 & 612.57 & 150.62 
& 58.31 & 2.55 & 40.84 & 15.27 \\

6 & 951.92 & 83.76 & 655.39 & 157.65 
& 55.12 & 2.43 & 40.96 & 14.64 \\

7 & 1021.93 & 80.68 & 740.52 & 149.59 
& 51.14 & 3.05 & 46.28 & 20.86 \\

8 & 943.33 & 81.05 & 662.22 & 147.51 
& 52.55 & 2.68 & 47.30 & 17.15 \\

9 & 1000.65 & 97.31 & 705.96 & 151.85 
& 45.53 & 2.81 & 50.42 & 18.53 \\

10 & 908.27 & 93.94 & 567.56 & 184.36 
& 62.41 & 2.60 & 31.53 & 12.79 \\
\midrule
\textbf{Average (s)}
& 936.72 & 84.64 & 649.55 & 150.45 & 52.08
& 2.72 & 43.95 & 17.12 \\
\bottomrule
\end{tabular}
\end{table*}


\begin{table*}[t]
\scriptsize
\centering
\caption{Coverage statistics across 10 sessions}
\label{tab:coverage-stats}
\footnotesize
\begin{tabular}{c|cccc|ccc|ccc}
\toprule
\multirow{2}{*}{\textbf{Sessions}} &
\multicolumn{4}{c|}{\textbf{MR Summary \& Requirement Coverage (\%)}} &
\multicolumn{3}{c|}{\textbf{Test Summary}} &
\multicolumn{3}{c}{\textbf{Mutation Coverage}} \\
\cmidrule(r){2-5} \cmidrule(l){6-8} \cmidrule(l){9-11}
& \textbf{Gen. MR's} & \textbf{Dropped} & \textbf{Refined} & \textbf{Coverage (\%)}
& \textbf{Generated} & \textbf{Passed (\%)} & \textbf{Failed (\%)}
& \textbf{Generated} & \textbf{Killed} & \textbf{Score} \\
\midrule
1  & 14 & 2 & 12 & 64.71 & 41 & 85.37 & 14.63 & 104 & 65 & 0.63 \\ 
2  & 16 & 1 & 15 & 35.29 & 60 & 76.67 & 23.33 & 176 & 83 & 0.47 \\ 
3  & 14 & 1 & 13 & 58.82 & 48 & 91.67 & 8.33  & 162 & 85 & 0.52 \\ 
4  & 17 & 1 & 16 & 70.59 & 55 & 83.64 & 16.36 & 174 & 83 & 0.48 \\ 
5  & 15 & 0 & 15 & 52.94 & 59 & 72.88 & 27.12 & 162 & 96 & 0.59 \\ 
6  & 16 & 0 & 16 & 82.35 & 65 & 63.08 & 36.92 & 94  & 63 & 0.67 \\ 
7  & 16 & 0 & 16 & 70.59 & 49 & 53.06 & 46.94 & 138 & 82 & 0.59 \\ 
8  & 15 & 1 & 14 & 64.71 & 55 & 70.91 & 29.09 & 158 & 91 & 0.58 \\ 
9  & 17 & 3 & 14 & 45.83 & 54 & 64.81 & 35.19 & 78  & 46 & 0.59 \\ 
10 & 18 & 0 & 18 & 58.82 & 71 & 78.87 & 21.13 & 126 & 57 & 0.45 \\ 
\midrule
\textbf{Average}
& 15.80 & 0.90 & 14.90 & 60.47 & 55.70 & 74.10 & 25.90 & 137.20 & 75.10 & 0.56 \\
\bottomrule
\end{tabular}
\end{table*}


\subsubsection{RQ1}
    
    AgenticMeta successfully executed the complete workflow, including extraction, MR generation, refinement, test generation, validation, instantiation, simulation, and mutation analysis in all ten sessions. 

The results show that the proposed architecture is feasible for metamorphic testing of FMU-based simulation models. Across all ten sessions, the workflow successfully progressed from requirement extraction to executable metamorphic test generation and execution under the specified configuration. It was also able to perform mutation-based quality assessment on top of the core execution.

\subsubsection{RQ2}

Across the 10 sessions, the workflow generated an average of 15.80 MRs and 
55.70 test cases per run.
The average requirement coverage was 60.47\%
(see Table \ref{tab:coverage-stats}, MR Summary \& Requirement Coverage). These results indicate that the workflow can repeatedly generate specification-linked MRs and executable metamorphic test cases based on the FMU interface.

    The results also show that the workflow can repeatedly produce specification-linked MRs and executable tests. The produced MRs were manually checked by a domain expert and considered to be meaningful and consistent with the system specification. However, full requirement coverage was not achieved because not all extracted requirements could be converted into valid MRs over the exposed FMU interface. Manual inspection suggested two main causes:
    
    \begin{itemize}[leftmargin=*]
        \item a few extracted requirements are generic statements about system behaviour and do not define a directly manipulable input-output relation. 
        \item a few requirements described output behaviours that cannot be directly influenced by the available FMU inputs. Such requirements were often removed during the MR refinement stage because the agent reasoned that a valid follow-up output could not be guaranteed with the given inputs.
    \end{itemize}

    It was observed that the MRs that were not covered or dropped were mainly those based on indirect, generic, or weakly controllable behaviours.  In the LOC case study, MRs based on direct relations between controllable inputs and observable outputs were more stable, while MRs requiring indirect influence over variables were more likely to be refined or removed. This indicates that the approach is most effective when the specification states explicit causal links between input perturbations and expected output responses.

    The test case pass rate across simulation sessions averaged 74.10\% 
    (see Table \ref{tab:coverage-stats}, Test Summary). This indicates that most generated tests were semantically aligned with the generated MRs. As the implementation of the LOC is considered to be correct, the failing tests in this case are considered false positives and a limitation of our workflow, which we will address in future work. 
    Further investigation showed that several failures occurred when the output variables initially moved in the expected direction but settled at a different value before satisfying the specified relation. Other failures occurred when the output settled outside the specified tolerance range or around a value different from the setpoint values. Both these situations can be due to the ramp-up values defined by the workflow. 
    These observations indicate that future versions of the workflow should include stronger validation of temporal operators and tolerance values.
  
\subsubsection{RQ3}
    The average mutation score across the ten sessions was 0.56, corresponding to 56\% of generated mutants killed (see Table \ref{tab:coverage-stats}, Mutation Coverage). These results indicate that the majority of mutations are detected by the evaluator, but they do not exhaustively cover the model behaviour.

    The mutation results should be interpreted cautiously. Because mutations are applied to recorded output time series rather than to the FMU implementation, the score reflects the sensitivity of the MRs to selected output deviations, not the full fault-detection capability of the workflow against implementation defects.
    In addition, the pass rate and mutation score do not always increase together. The reason was that a higher pass rate means more test cases satisfy their original MRs, and mutation analysis is then applied to a larger set of passing tests. This increases the number of generated mutants. However, some mutation operators 
    are less effective for certain relation types. In these cases, the mutated outputs may still satisfy the relation, so the mutants are not killed. Consequently, sessions with more passing tests may generate more surviving mutants, which reduces the overall mutation score even when the test pass rate is high.

\subsubsection{RQ4}

    The average test-generation time is 2.72 seconds per test case, 
    the average MR-generation and refinement time is 43.95 seconds per accepted MR, \& 
    the average execution time per test case is 17.12 seconds
    (see Table \ref{tab:runtime_stats}, Unit level statistics). These results show that the overall time taken for the complete loop from extraction to execution is very low per test case. The compact test-case schema generated by the test generator agent is effective in keeping the test generation phase relatively fast.

The main efficiency bottleneck is the LLM-based MR generation and refinement. The refinement phase remains comparatively expensive due to the 'high' reasoning effort of \textit{MR Refiner Agent}, and also it checks generated MRs against both structured extraction output and the original functional specification, making it a resource-intensive process.
The result suggests that future optimization should focus on reducing unnecessary MR refinement calls, prioritizing uncovered requirements more selectively, caching stable extraction artifacts, and using less expensive models or reasoning settings for low-risk validation steps.

\section{Threats to Validity}
\label{sec:limitations}

In this section, we discuss the possible threats to the validity of our study. First, the evaluation is conducted on a single FMU-based case study, the LOC system. Although LOC is representative of dynamic simulation models with continuous inputs and outputs, the findings may not generalize to models with substantially different control logic, interface structures, or operating characteristics.

Second, the workflow depends on the quality of the functional specification. As the implementation uses document conversion tools and evidence-preserving extraction prompts, the generated MRs are directly constrained on the quality of the specification document. In addition, the workflow cannot reliably generate MRs for variables that are not described in \emph{model\_description.xml}, since the model interface description is treated as the authoritative source for available FMU inputs and outputs.

\section{Conclusions \& Future Work}
This paper proposed AgenticMeta, a multi-agent workflow for specification-based metamorphic testing of FMU-based simulation models. The approach used functional and interface specifications to extract test-relevant requirements, generate and refine MRs, produce executable metamorphic test cases, and evaluate test outcomes using metamorphic oracles and mutation analysis. The feasibility evaluation was performed on the LOC system, which shows that the workflow can stably execute the complete testing across repeated sessions. AgenticMeta generated specification-linked MRs and executable test cases, achieved an average requirement coverage of 60.47\%, an average test-case pass rate of 74.10\%, and an average mutation score of 56\%. These results indicate that the approach is feasible for supporting automated metamorphic testing of FMU-based simulations. At the same time, the results also highlight important limitations. Requirement coverage is constrained by the quality and testability of the specification.

Future work will extend the workflow by transforming the current deterministic \textit{Coordinator node} into an LLM-based agent. This would allow the Coordinator to make adaptive decisions during execution, such as prioritizing uncovered requirements, deciding when additional MR refinement is needed, selecting alternative strategies, and determining whether the workflow should continue or stop based on coverage and test-quality feedback. In addition, future work will aim to improve mutation analysis by designing mutation operators that are more closely aligned with the semantics of the evaluated MRs. This would make mutation-based assessment more robust for FMU systems.

\section*{Acknowledgments}
This work was funded by the Finnish Ministry of Education and Culture's Doctoral Education Pilot under Decision No. VN/3137/2024-OKM-6 (The Finnish Doctoral Program Network in Artificial Intelligence, AI-DOC) and Business Finland via the Virtual Sea Trial project (VST), under grant 7187/31/2023.

\end{document}